# Extensivity and Relativistic Thermodynamics

**J. Dunning-Davies,**
**Department of Physics,**
**University of Hull,**
**Hull HU6 7RX.**

email:   j.dunning-davies@hull.ac.uk

**Abstract.**

The mathematical properties associated with the widely accepted concept of the extensivity of many of the common thermodynamic variables are examined and some of their consequences considered. The possible conflict between some of these and currently accepted results of special relativistic thermodynamics is highlighted. Although several questions are raised, answers are not advanced as this seems an area demanding calm, widespread reflection which could conceivably lead to radical revision of part, or parts, of theoretical physics.

**Introduction.**

In his thermodynamics book of 1961 [1], Landsberg emphasised the importance of the concept of extensivity in the development of several aspects of the subject. He went so far as to suggest that the assumption that all thermodynamic variables were either extensive or intensive shared some important characteristics with the 'laws' of thermodynamics and this, in turn, provoked him to suggest that it might be appropriate to regard that assumption as a 'fourth law of thermodynamics'. In a footnote, he commented that the need for new assumptions, or postulates, is often not stressed enough but he cited, as a notable exception, Prigogine's book [2], *Thermodynamics of Irreversible Processes*, where, on page 93, it is clearly stated that the Gibbs' formula
$$TdS = dU + pdV - \mu dN$$
'was originally proved for the equilibrium conditions, and its use for the non-equilibrium conditions is a new postulate on which the whole of the thermodynamics of irreversible processes is based'. Landsberg then used this 'fourth' law to deduce those two formulae so vitally important in thermodynamics:-
the Euler relation
$$TS = U + pV - \mu N$$
and the Gibbs-Duhem relation
$$SdT - Vdp + Nd\mu = 0,$$
which shows that the intensive variables are not all independent.

The whole notion of extensivity was examined in a series of three articles [3] and it emerged that the actual mathematical property of interest and relevance here is that of homogeneity of degree one. All the necessary mathematical details are covered in Gillespie's short text [4] in which a function $f(x, y, z)$ is said to be homogeneous of degree one in the variables $x, y, z$ if
$$f(ax, ay, az) = af(x, y, z)$$
Euler's theorem on homogeneous functions, which states that, if $f(x, y, z)$ is a homogeneous function of degree one in $x, y, z$ then
$$x\frac{\partial f}{\partial x} + y\frac{\partial f}{\partial y} + z\frac{\partial f}{\partial z} = f,$$
is then proved, as is its converse. This latter point might be noted since, all too often, Euler's theorem is stressed with little or no mention made of the fact that the converse holds also – or, mathematically, the proof is an 'if and only if' one. This seemingly small point proves useful in the context of thermodynamics, as will be shown.

**Extensivity in thermodynamics.**

The concept of extensivity in thermodynamics is used when, as is shown quite clearly in Landsberg's book [1], extending ideas to cover open and non-equilibrium systems. Up to this point in the development of the subject, it has been assumed already that the heat and, therefore, the entropy are additive functions and, as has been shown previously [3], this amounts to the entropy being an extensive function of its variables. This fact, together with the well-known properties of homogeneous functions, are used to derive the important and well-known Euler relation linking the extensive and intensive thermodynamic variables
$$TS = U + pV - \mu N.$$

As mentioned above, this equation is then used in conjunction with the equation reflecting the combination of the first and second laws -

$$TdS = dU + pdV - \mu dN$$

- to deduce the Gibbs-Duhem relation

$$SdT - Vdp + Nd\mu = 0.$$

Hence, it is seen that the assumption of extensivity of the entropy leads directly to several important and useful thermodynamic relations. It should be noted that lack of extensivity not only rules out use of these two equations but also other relations whose derivation depends on this property. A typical example of this is provided by derivation of the expression for the mean square relative fluctuation in number of particles which is in terms of the isothermal compressibility $K_T$:

$$\frac{\overline{(N-\overline{N})^2}}{(\overline{N})^2} = -\frac{kT}{V^2}\frac{1}{(\partial P/\partial V)_{N,T}} = \frac{k}{T}K_T.$$

As has been shown [5], the derivation of this result depends crucially on two facts: the validity of the Gibbs-Duhem relation and the pressure being a function of both the volume and the temperature. It follows that this particular relation does not hold when considering fluctuations below the condensation temperature of a system since, below that temperature, the pressure becomes independent of the volume.

However, it might be noted also that, if

$$TS = U + pV - \mu N$$

is assumed valid and if the usual definitions for $T$, $p/T$ and $\mu/T$ -

$$\frac{1}{T} = \left(\frac{\partial S}{\partial U}\right)_{V,N}, \quad \frac{p}{T} = \left(\frac{\partial S}{\partial V}\right)_{N,U}, \quad \frac{\mu}{T} = -\left(\frac{\partial S}{\partial N}\right)_{U,V}$$

- are used, the converse of Euler's theorem shows that the entropy, $S$, must be a homogeneous function of $U$, $V$ and $N$. Hence, whatever argument is adopted to derive the Euler relation, homogeneity of the entropy is assumed either directly or indirectly.

Further, it might be noted that, if the functions involved are mathematically well-behaved and, in general, functions in physics do satisfy this requirement, then the equation

$$S = S(U, V, N)$$

may be solved to give

$$V = V(S, U, N)$$

and, therefore, the variables $V$, $U$ and $N$ are also seen to be extensive. This is all in accordance with what has been written and accepted for many years [1]. However, it does not appear to have been noticed that this does not fit in with several widely accepted results associated with special relativistic thermodynamics.

**Extensivity and relativistic thermodynamics.**

Once the results of special relativity are considered, the first consequence for thermodynamics is that, under the Lorentz transformation, volume will transform according to

$$V \rightarrow \beta^{-1}V$$

or, written out in more detail to bring out the thermodynamic consequences

$$V(S, U, N) \rightarrow \beta^{-1}V(S, U, N),$$

where $\beta = (1 - v^2/c^2)^{-1/2}$.

However, according to earlier discussions, the volume, *V*, is an extensive variable; that is, it is homogeneous of degree one and, as a consequence, the above relativistic transformation law would appear to imply

$$V(S, U, N) \rightarrow \beta^{-1}V(S, U, N) = V(\beta^{-1}S, \beta^{-1}U, \beta^{-1}N)$$

and, again due to the property of extensivity, this would appear to imply obvious transformation laws for the other extensive thermodynamic variables, transformation laws which are certainly not in agreement with presently accepted practice. Firstly, entropy is normally taken to be invariant under Lorentz transformation and it is difficult to see how *N*, the number of particles, would change under a mathematical transformation but, here, the clear implication is

$$S \rightarrow \beta^{-1}S \quad \text{and} \quad N \rightarrow \beta^{-1}N.$$

A further consequence, if these transformation equations are used, is that the intensive variables will all be invariant under Lorentz transformation. This would certainly clear up the controversy surrounding the question of the transformation of temperature and would retain the notion of invariance for pressure, but at what price?

An area which needs further consideration is that concerning the transformation of the internal energy, denoted here by *U*. Tolman [6] does not consider the internal energy alone but examines rather the supposed behaviour of (*U* + *pV*). He finds that this quantity transforms according to

$$U + pV \rightarrow \beta(U + pV)$$

and so, it follows that *U* transforms according to

$$U \rightarrow \beta(U + pV) - \beta^{-1}pV = \beta\left(U + \frac{v^2}{c^2}pV\right)$$

However, it is claimed also that temperature transforms according to

$$T \rightarrow \beta^{-1}T$$

Hence, if *S* is invariant under Lorentz transformation, (*U* + *pV*) and *TS* do not transform in the same way and so, the above Euler relation cannot hold. The same conclusion also follows from the approach adopted by Pauli [7], as is seen quite clearly from section 46 of the referenced text. Hence, in both these classic references, the Euler relation is seen not to hold in the transformed frame, but in neither case does this point seem to have been noted. However, from the manner of writing, it appears that the quantity represented by the letter *E* in both the writings of Tolman and Pauli, which at first sight should be analogous with the internal energy, *U*, considered here, is actually something else. In Tolman's case, it is claimed to be energy content and, in Pauli's, it is referred to as energy but there seems some indication that by this is meant kinetic energy. However, as has been very clearly shown by Chandrasekhar [8], the kinetic energy is definitely not the same as the internal energy, - at least, not in general. Chandrasekhar shows that, in the case of a perfect gas, the kinetic and internal energies are the same only when the ratio of the heat capacities has the value 5/3.

**Conclusions.**

This entire discussion means that the question raised is either

'Is extensivity in thermodynamics compatible with special relativity?'

or

'Is special relativity compatible with extensivity in thermodynamics?'

If the answer to one or other of these questions is 'No', a radical review of special relativistic thermodynamics will be necessary. It will be vital to ascertain precisely which familiar results remain valid in relativistic circumstances and which do not. This, in itself, need not be a disastrous situation. It would, indeed, be analogous to that now facing people who accept the somewhat dubious validity of the so-called Bekenstein-Hawking expression for the entropy of a black hole. That expression is not extensive and so there can, for example, be no Euler relation or Gibbs-Duhem equation applicable. Of course, the stated entropy expression is also not concave and so grave doubts over its actual validity must remain. However, as far as the question of extensivity in thermodynamics and special relativity is concerned it might be remembered that thermodynamics rests on a relatively firm experimental /observational foundation. The same cannot be said for several sections of special relativity, particularly that part totally reliant on mathematical reasoning.

Thermodynamics and its laws, which are really facts of experience, have been successful in helping to explain so many and such varied physical phenomena, from the microscopic to the macroscopic, for many years. Special relativity has had much success also but over a somewhat shorter period of time. This note merely highlights a possible problem within the intersection of these two theories. This point gains added significance when the position of the revolutionary new Hadronic Mechanics of Santilli [9] is considered. Santilli's starting point is simply to recognise the need for new mathematics when attempting to deal with some of the outstanding problems facing modern science. The interesting point in the present context is that, although these new mathematical formulations of Hadronic Mechanics appear radically different from the more orthodox formulations with which everyone is so familiar, the thermodynamic formulæ remain unaltered [10].

The discussion also raises, once again, the possible problem associated with the notion within thermodynamics of internal energy. As clearly pointed out by Chandrasekhar, this is not the same as kinetic energy in general and this seemingly trivial point certainly seems to require emphasis when introducing thermodynamics initially and should be borne in mind constantly by all when making use of thermodynamic results.

Again, it might be noted that, if extensivity is removed as an important property of thermodynamic variables, especially the entropy, then it must be remembered that the mathematical property of concavity of the entropy becomes of paramount importance [11]. If this property is ignored, as has been seen to be the case with the Bekenstein-Hawking expression for the entropy of a black hole, dire consequences can follow. In particular, negative heat capacities in closed systems would occur and this could result in violations of the Second Law of Thermodynamics. In systems for which the entropy is merely concave but not extensive, relations such as those due to Euler and Gibbs-Duhem would be valid no longer. Hence, it is really the concavity property of the entropy which is vitally important in ensuring the validity of the second law but classical thermodynamics relies heavily on relations such as these latter two in many manipulations and it should be remembered that it is doubtful anyone has followed through *all* the consequences of the entropy being merely concave but not extensive.

Finally, it should be remembered that this is not the first time queries have been raised concerning problems associated with relativistic thermodynamics. Serious problems in this area were highlighted when errors associated with the inflationary scenario were discussed [12]. It was clearly shown then that, when dissipative processes are considered, the condition for the vanishing of the divergence of the mass-energy tensor is a combination of the first and second laws, rather than being the relativistic analogue of the first law, as had been claimed. Since the first and second laws are concerned with totally different physical entities, it seems incorrect that a mathematical equation should be able to combine them. Physical reality would certainly appear to indicate that these two powerful laws are independent of each other; both are necessary in their own right in the development of thermodynamics as a subject. Hence, queries are immediately raised about the relevant condition which seems, in one sense, to combine them. It was shown also in the quoted article that dissipative processes destroy isotropy and hence may not be described by any standard model based on Robertson-Walker models. This latter point is yet another to be born in mind when attempts are made to combine relativity with thermodynamics.

**References.**

[1] P. T. Landsberg; 1961, *Thermodynamics with Quantum Statistical Illustrations*,
(Interscience, New York)

[2] I. Prigogine;1967, *Thermodynamics of Irreversible Processes*,
(Interscience, New York)

[3] J. Dunning-Davies; 1983, Physics Letters, **94A**, 346; **97A**, 327
J. Dunning-Davies & P. T. Landsberg; 1985, Physics Letters, **107A**, 383

[4] R.P.Gillespie; 1954, *Partial Differentiation*,
(Oliver and Boyd, Edinburgh)

[5] J. Dunning-Davies; 1968, Il Nuovo Cimento, **53B**, 180; **57B**, 315

[6] R.Tolman; 1934, *Relativity, Thermodynamics and Cosmology*,
(Oxford UP, Oxford)

[7] W. Pauli; 1958, *Theory of Relativity*,
(Pergamon Press, Oxford)

[8] S. Chandrasekhar; 1957, *An Introduction to the Study of Stellar Structure*,
(Dover, New York)

[9] R. M. Santilli; 2001, *Foundations of Hadronic Chemistry*,
(Kluwer Academic Publishers, Dordrecht)

[10] J. Dunning-Davies; 1999, Found. Phys. Lett, **12**, 593
2006, Progress in Physics, **4**, 24

[11] B. H. Lavenda & J. Dunning-Davies; 1990, Found. Phys. Lett. **3**, 435
J. Dunning-Davies; 1993, Found. Phys. Lett. **6**, 289
1994, Trends in Stat. Phys. **1**, 233

[12] B. H. Lavenda & J. Dunning-Davies; 1992, Found. Phys. Lett. **5**, 191